\documentclass[ reprint,
 amsmath,amssymb,
 aps,
 pre,
]{revtex4-1}
\usepackage{graphicx}
\usepackage{dcolumn}
\usepackage{bm}
\usepackage{float}
\usepackage{xcolor}
\usepackage{subfigure}
\usepackage[abs]{overpic}
\usepackage{hyperref}

\begin{document}
\title{Target search on DNA by interacting molecules: First-passage approach}
\author{Jaeoh Shin$^{1,2}$}
\author{Anatoly B. Kolomeisky$^{1,2,3,4}$}

\affiliation{$^1$Department of Chemistry, Rice University, Houston, Texas, 77005, USA}
\affiliation{$^2$Center for Theoretical Biological Physics, Rice University, Houston, Texas, 77005, USA}
\affiliation{$^3$Department of Chemical and Biomolecular Engineering, Rice University, Houston, Texas, 77005, USA}
\affiliation{$^4$Department of Physics and Astronomy, Rice University, Houston, Texas, 77005, USA}

\date{\today}

\begin{abstract}

Gene regulation is one of the most important fundamental biological processes in living cells. It involves multiple protein molecules that locate specific sites on DNA and assemble gene initiation or gene repression multi-molecular complexes. While the protein search dynamics for DNA targets has been intensively investigated, the role of inter-molecular interactions during the genetic activation or repression remains not well quantified. Here we present a simple one-dimensional model of target search for two interacting molecules that can reversibly form a dimer molecular complex, which also participates in the search process. In addition, the proteins  have  finite residence times on specific target sites, and the gene is activated or repressed when both proteins are simultaneously present at the target. The model is analyzed using first-passage analytical calculations and Monte Carlo computer simulations. It is shown that the search dynamics exhibits a complex behavior depending on the strength of inter-molecular interactions and on the target residence times. We also found that the search time shows a non-monotonic behavior as a function of the dissociation rate for the molecular complex. Physical-chemical arguments to explain these observations are presented. Our theoretical approach highlights the importance of molecular interactions in the complex process of gene activation/repression by multiple transcription factor proteins.

\end{abstract}

\maketitle

\section{Introduction}

Gene regulation is a crucial biological process that supports the successful functioning of all living systems \cite{lodish2008molecular,alberts2013essential,phillips2012physical,lelli2012disentangling}. Responding to external environmental signals or to signals from other biological cells, several classes of protein molecules, known as transcription factors, are activated. These proteins are assembled then at specific regulation regions on DNA by forming multi-molecular complexes, which activates or represses the specific genes by respectively increasing or decreasing the level of transcription \cite{lodish2008molecular,alberts2013essential}. A significant progress in our understanding of how genes are turned on or turned off in cells has been achieved in recent years \cite{lelli2012disentangling}. However, despite its critical importance for the survival of biological cells, the molecular mechanisms of gene regulation still remain not fully understood \cite{coulon2013eukaryotic,phillips2012physical,lelli2012disentangling}.

The initial stage of gene regulation is the process of transcription factors searching for specific sequences on DNA. It has been intensively investigated in the last 40 years using various approaches \cite{riggs1970lac,berg1976association,mirny2009,halford2004,coppey2004,sheinman2012classes,kolomeisky2011,hu2006proteins,veksler2013speed,shvets2018mechanisms,shin2018a,bauer2012generalized,hammar2012lac}. Experimental studies suggest that in the target search the protein molecules experience both three-dimensional (in the bulk solution) and one-dimensional motions (along the DNA chain), which leads to unexpectedly high protein-DNA effective association rates in some systems. This is known as a {\it facilitated diffusion} phenomenon \cite{mirny2009,halford2004,kolomeisky2011}. Multiple theoretical ideas to explain these observations have been proposed, emphasizing the role of non-specific protein-DNA interactions, conformational transitions and inter-segment transfer for protein molecules bound non-specifically on DNA \cite{shvets2018mechanisms}.

Although theoretical investigations of protein search dynamics have  clarified many aspects of biochemical and biophysical phenomena during the early stages  of gene regulation, one important aspect of these processes is not taken into account in these studies. Participating protein molecules interact with each other, and the gene activation or repression will start only after multi-molecular protein complexes are fully assembled at the target sites. It is known that the pre-initiation complexes that are necessary for the transcription of genes typically contain more than 100 protein molecules of different types \cite{lodish2008molecular,alberts2013essential}. A recent experimental study also specifically showed that tuning the degree of polymerization in transcription factors Yan strongly modifies gene repression in {\it Drosophila}, underlying the importance of protein-protein interactions \cite{hope2018tuned}. Generally, proteins can produce a wide spectrum of conformational states and multi-molecular complexes that might strongly influence the levels of genetic regulation in cells \cite{siggers2013protein}.

There are only few theoretical studies that address the role of inter-molecular couplings during the target search \cite{grebenkov2017first,lawley2019first}. Mean first-passage times (MFPT) for two independent particles with reversible target-binding kinetics to reach simultaneously the target in a one-dimensional system have been analytically calculated in ref. \cite{grebenkov2017first}. However, in this work, the particles do not directly interact with each other, and they are only coupled indirectly due to finite residence times on the target site for each particle. The method utilizes a continuum diffusion on the interval with the target located at one end of the interval. This approach was recently generalized for many particles in various dimensions \cite{lawley2019first}, and approximate expressions for first-passage time distributions (which are exact in certain limiting cases) have been obtained. However, there are no theoretical studies that take into account explicitly the inter-molecular interactions.

In this paper, we present a minimal theoretical model that investigates the target search dynamics of two {\it interacting} particles on a one-dimensional lattice with arbitrary position of the target and with reversible target-binding kinetics. The molecules can form a dimer complex that can dissociate back into separate particles, and the search process ends when two particles are found simultaneously at the target site. The dynamics in the system is analyzed in several limiting cases using a first-passage approach, which was successful in studies of multiple processes in chemistry, physics and biology \cite{redner2001guide, benichou2014first, metzler2014first,kolomeisky2015motor}. We also investigate the search dynamics at general conditions employing extensive Monte Carlo computer simulations. Even though the model is rather simple, it shows a rich dynamic behavior depending on the values of transition rates and diffusion coefficients of the searching particles. We present physical-chemical arguments to explain the complex dynamics in the system. Our theoretical method clarifies the role of inter-molecular interactions in the target search phenomena.

\section{Model}
\label{sec-model}

\begin{figure}
    \centering
    \includegraphics[width=0.9\columnwidth]{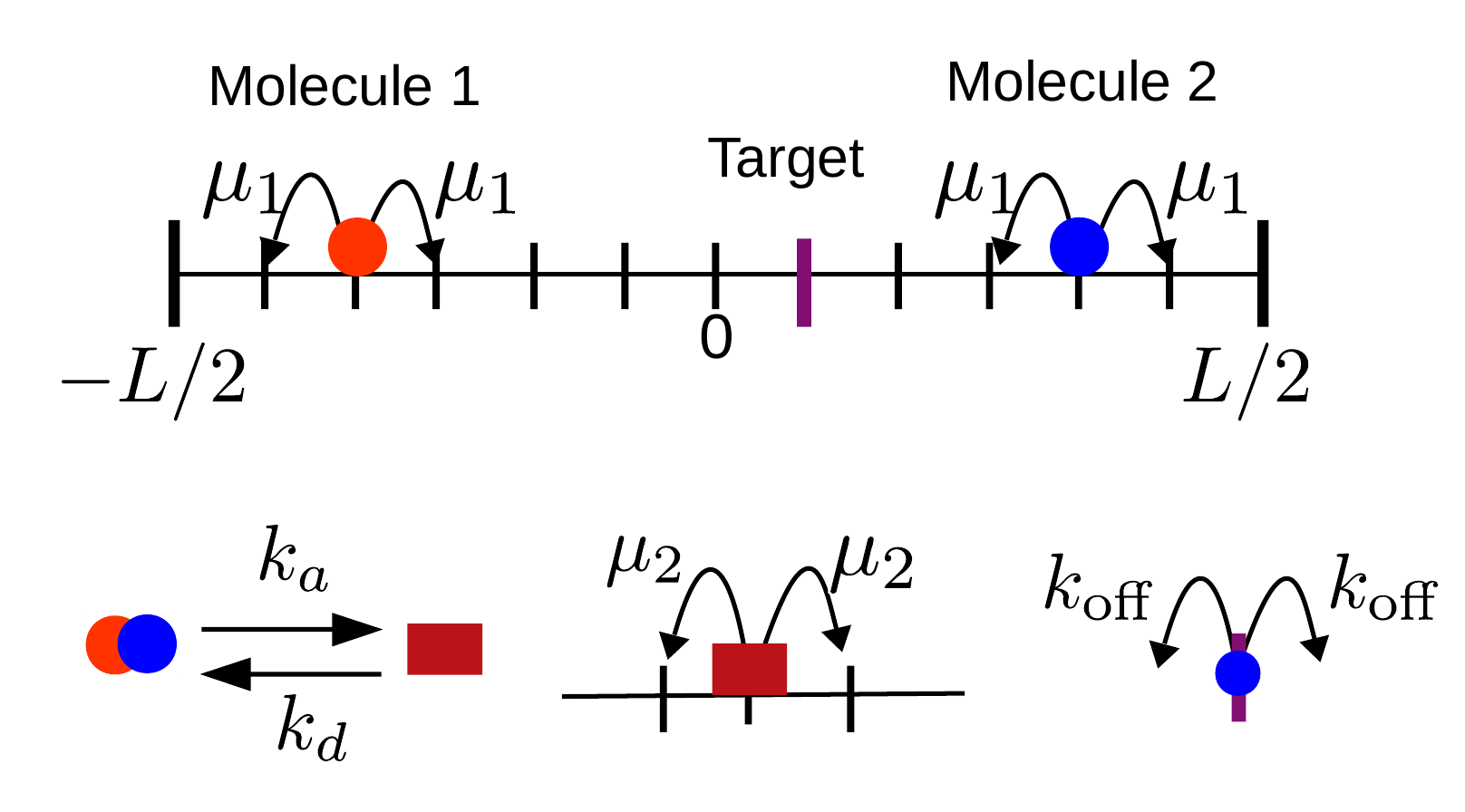}
    \caption{A schematic view of the model. Two molecules (red and blue circles) are moving on the lattice of size $L+1$ with the hopping rate $\mu_1$. When two molecules occupy the same site, they can form a complex with the association rate $k_a$ and dissociate from the complex with the rate of $k_d$. The complex moves on the lattice with the hopping rate of $\mu_2$. When the molecule arrives at the target site it can  leave the site with the rate $k_{\text{off}}$. }
    \label{fig-1}
\end{figure}

Let us consider a system shown in Fig. \ref{fig-1}. There are two molecules that diffuse along a one-dimensional lattice (which corresponds to a DNA chain) with a hopping rate $\mu_{1}$ in both directions. The lattice has $L+1$ sites ($L$ is even number), which are labeled as $-L/2$, $-L/2+1$, ... , and $+L/2$, and one of them is a special target site: see Fig. \ref{fig-1}. The two molecules might form a complex with a rate $k_{a}$ when they are found at the same site. The complex can diffuse along the lattice with a hopping rate $\mu_{2}$, and it also can dissociate back into separate particles with a rate $k_{d}$ (Fig. \ref{fig-1}). The association and the dissociation rates are related via a detailed-balance like expression,
$\frac{k_a}{k_d}=\exp(-\frac{E}{k_{\text B}T})$, 
where $E$ is the interaction energy between two molecules. It has the following physical meaning. The stronger the attractive interaction ($E \ll -1 k_{B}T$), the more probable is the formation of the complex. But increasing the repulsion ($E \gg 1 k_{\text{B}}T$) lowers the probability of the particles association. We assume that the particles interact only when they are found at the same site. For the case when the association rate is much faster than the diffusion ($k_a \gg \mu_1$), the complex formation is almost instantaneous once two molecules arrive at the same site. 
In the opposite limit of slow association or fast dissociation rates ($k_a \ll \mu_1$ or $\mu_1 \ll k_{d}$), two molecules cannot make the complex, or the produced complex immediately breaks apart, and hence the particles move independently. Because the diffusion rates reflect the effective interactions between the lattice (the DNA chain) and the protein molecules or complexes, we generally expect that $\mu_{2} < \mu_{1}$.

In our model, the interaction between particles and the lattice is different at the special target site. The particle have finite residence times at the target: see Fig. \ref{fig-1}. The molecule can leave this site with an unbinding rate $k_{\text{off}}$ due to the finite molecule-target interactions. It is assumed that the search process ends when two molecules arrive, either individually or as the complex, at the target site simultaneously. When the particles reach the target site as the dimer complex, the process will end immediately. However, when the molecules reach the target site as  monomers, the finite residence times strongly modify the dynamics in the system, as was already discussed earlier \cite{grebenkov2017first}.

\subsection*{Simulation method}

We perform kinetic Monte-Carlo simulations of the model in Fig. \ref{fig-1} using a method described in ref. \cite{kafri2005sequence}. Our model has total five kinetic parameters: $\mu_1$, $\mu_2$,  $k_a$, $k_d$ and $k_{\text{off}}$. We set $\mu_1=0.25$, and all other parameters are normalized to that value. The time of the system increases as $t_{i}=\Delta t \times i$, where $\Delta t={(4 \mu_1)}^{-1}$ and $i=0, 1, 2, ...$ . We use the time in the units of $\Delta t$ in the following. When the molecule is in the monomer state, it has three possibilities in the next step: it remains at the same site with the probability $1/2$, or it can hop onto its right or left sites with the probabilities $1/4$ in each direction, respectively. If the molecules are at the boundary of the lattice ($n=\pm L/2$), it can move only to the available site away from the boundary. Considering the molecular complex, there are also three possibilities in the next step. (i) It dissociates into two molecules with the probability $k_d \Delta t$;(ii) it moves as the dimer with the probability $\mu_2 \Delta t$ to the neighboring site; (iii) it remains at the same site with the probability $\max \{0,  (1-k_d \Delta t -\mu_2\Delta t) \}$. The dissociation event is modeled as the process in which one of the molecules from the complex is hopping to the neighboring site, while the other molecule stays at the same location. Initially, the two molecules are randomly distributed on the lattice with the equal probabilities. Our main interest is the time for the two molecules to arrive at the target site located at the site $m$ for the first time.

\section{Results}
\label{sec-results}

To understand better the mechanisms of the search phenomena in the system of interacting particles, we analyze first several limiting situations for which analytical results can be obtained.

\subsection{Strong repulsions: non-interacting molecules ($k_a \ll \mu_1$ or $\mu_1 \ll k_{d}$)}

We start with analyzing a non-interacting limit when the complex formation is very slow or the dissociation is very fast, i.e., for $k_a \ll \mu_1$ or $\mu_1\ll k_{d} $. This corresponds to the situation when particles strongly repel each other. As two molecules do not have time to form the complex even when they are at the same site, essentially all the time they move independently. In this case, the search dynamics depends on the particle residence time $1/k_{\text{off}}$ at the target site \cite{grebenkov2017first}. When $k_{\text{off}}=0$, the first molecule reaches the target site and it will remain there forever. The overall search process ends when the second particle arrives at the target. 

Let us consider a general case of $N$ non-interacting particles searching for the target site. The distribution of first passage times for $N$ molecules can be written as $f(N,t)=-\frac{\partial S(N,t)}{\partial t}$, where $S(N,t)$ is the probability that any of the $N$ molecules does not reach yet the target site at time $t$. This is known as a survival probability, and it can be written as
\begin{equation}
S(N,t)=1- \prod_{i=1}^{N} \left[1-S_i(1, t)\right],
\end{equation}
where $S_{i}(1,t)$ is the survival probability for the $i$-th molecule. Then the MFPT to activate the target for the system of $N$ molecules can be found from
\begin{equation}
t_N=\int_{t=0}^{\infty} t f(N,t)dt=-\int_{t=0}^{\infty} t \frac{\partial S(N,t)}{\partial t}dt=\int_{t=0}^{\infty} S(N,t) dt.
\end{equation}
Our main assumption here is that the survival probability for the single particle can be approximated as an exponential function
\begin{equation}
    S(1, t) \simeq \exp(-t/t_1),
\end{equation}
where $t_{1}$ is the MFPT for the single particle. This leads to
\begin{equation}
t_N \approx \int_{t=0}^{\infty}[ 1- (1-\exp(-t/t_1))^N ] dt,
\end{equation}
from which the mean search time for the system of $N$ particles can be always easily evaluated. For example, we  have $t_2=1.5 t_1$ for two independent particles and $t_3=\frac{11}{6}t_1$ for three independent particles. 

\begin{figure}
    \centering
    \includegraphics[width=0.9\columnwidth]{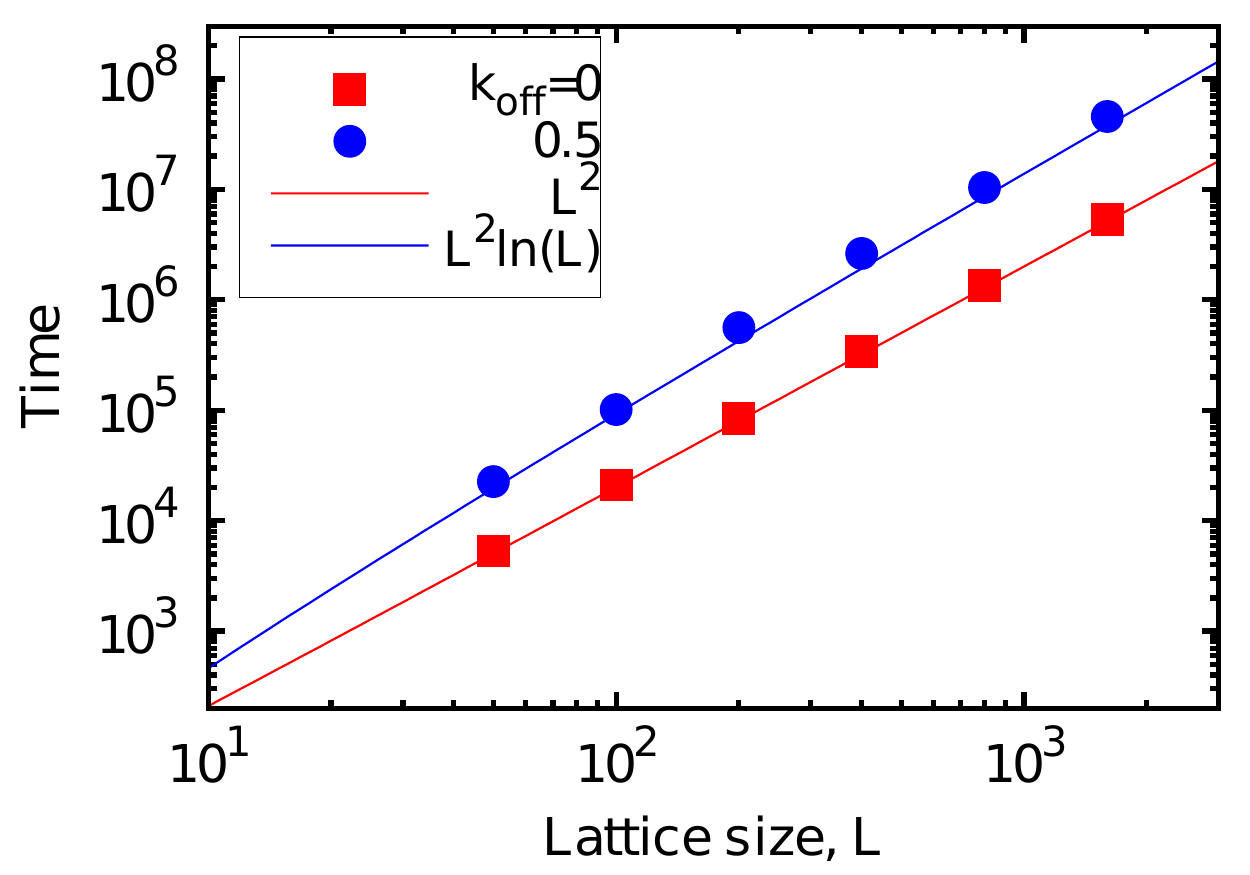}
    \caption{The target search time as a function of the lattice size $L$ for the irreversible ($k_{\text{off}}=0$) and reversible binding ($k_{\text{off}}=0.25$) to the target site in the strong repulsion limit when the dimer complex does not form. Here the target is located at the right end of the lattice $n=L/2$. }
    \label{fig-2}
\end{figure}

Returning to our system with two non-interacting molecules ($N=2$), it is known that for the 1D random walk \cite{veksler2013speed},
\begin{equation}
  t_1=\frac{2L^2+L}{6\mu_1},  
\end{equation}
yielding for the MFPT of two particles,
\begin{equation}
t_2=\frac{2L^2+L}{4\mu_1}.   
\end{equation}
The results of our theoretical predictions as compared to computer simulations are presented in Fig. \ref{fig-2}, and excellent agreement is found. The mean search time increases as $t_{2}=T(2,L)\propto L^2$, indicating that the search process is effectively the one-dimensional process, in agreement with our theoretical arguments. This is because it corresponds to the search time for the particle that arrives last to the target site.

When the residence time at the target site is finite, $k_{\text{off}}>0$, the dynamics of two molecules becomes more complicated \cite{grebenkov2017first}. For the special case of $k_{\text{off}}=\mu_1$ (=0.25), the target site behaves exactly the same as a reflecting boundary. 
Then the system of two independent searching particles on 1D lattice is analogous to a {\it single} random walker searching for a target on 2D lattice. The MFPT for such systems has been evaluated explicitly \cite{condamin2005first, condamin2007random}. In this case, the target search times grow as $T(2,L)\propto L^2 \ln (L)$, and this fully agrees with the results of Monte Carlo computer simulations (see Fig \ref{fig-2}).
\subsection{Strong attractions: instantaneous complex formation ($k_d \ll \mu_1 \ll k_{a}$)}

Another important limit is the case of the strong interaction when the two molecules form the dimer complex instantaneously once they arrive at the same site and the complex is stable for very long times ($k_d \ll \mu_1 \ll k_{a}$).  This corresponds to the case of strong inter-molecular attractions. The overall search process in this case consists of two steps. Two individual molecules must first come to the same site where the dimer complex is immediately formed, and after that, the search is taking place only by the complex. This suggests that the overall search time can be written as
\begin{equation} \label{T_complex}
    T=T_{enc}+T_{1D,complex},
\end{equation}
where $T_{enc}$ is the first encounter time of two molecules, and $T_{1D,complex}$ is the target search time for the complex. Since the hopping rate of the complex is generally slower than that for the monomer particles, the second term in Eq. (\ref{T_complex}) is dominating. The initial encounter can take place at any location, and this leads to the following approximation,
\begin{equation}
    T_{1D,complex} \simeq \frac{1}{L+1} \sum_{-L/2}^{L/2} T_{n},
\end{equation}
where we have \cite{veksler2013speed},
\begin{equation}
    T_{n}=\frac{(L/2-n)(3L/2+n+1)}{2\mu_2}.
\end{equation}
The final expression for the MFPT is given by
\begin{equation}
 T \approx \frac{1}{6\mu_2} (2L^2+L).  
\end{equation}
This formula approximates well the overall search time as indicated in the simulation results: see Fig. {\ref{fig-4} for small dissociation rates. One can also see that increasing the hopping rate for the complex lowers the overall search time, as one would expect.

\subsection{General situation}

Now let us consider a general situation when the target search is accomplished by both the monomer particles and the dimer complex. The results of our calculations and Monte Carlo computer simulations are presented in Fig. \ref{fig-4} where the MFPT are shown as a function of the dissociation rate $k_{d}$ for different values of the complex hopping rates $\mu_{2}$. Three dynamic search phases can be identified. When the dissociation rate is very large so that the complex does not contribute to the search we have an effective 2D dynamic phase since two articles move independently in the system. In this case, analytical calculations describe perfectly the search dynamics, as was already discussed above. In the opposite limit of very low dissociation rates, the complex forms very quickly and the search dynamics is effectively 1D because only one particle, the complex, is doing the search at these conditions. Again, we have a very good analytical description of this dynamic phase.  For intermediate values of inter-molecular interactions, the search dynamics can be viewed as a combination of 1D and 2D regimes because both the monomer particles and the dimer complex participate in the search process. Surprisingly, it is found that in this regime the search time  might be the fastest, and there is a non-monotonic dependence of the MFPT as a function of the dissociation rate $k_d$. This implies that there is an optimal inter-molecular interaction that might accelerate the overall target search. This is one of our main results. 
\begin{figure}
    \centering
    \includegraphics[width=0.9\columnwidth]{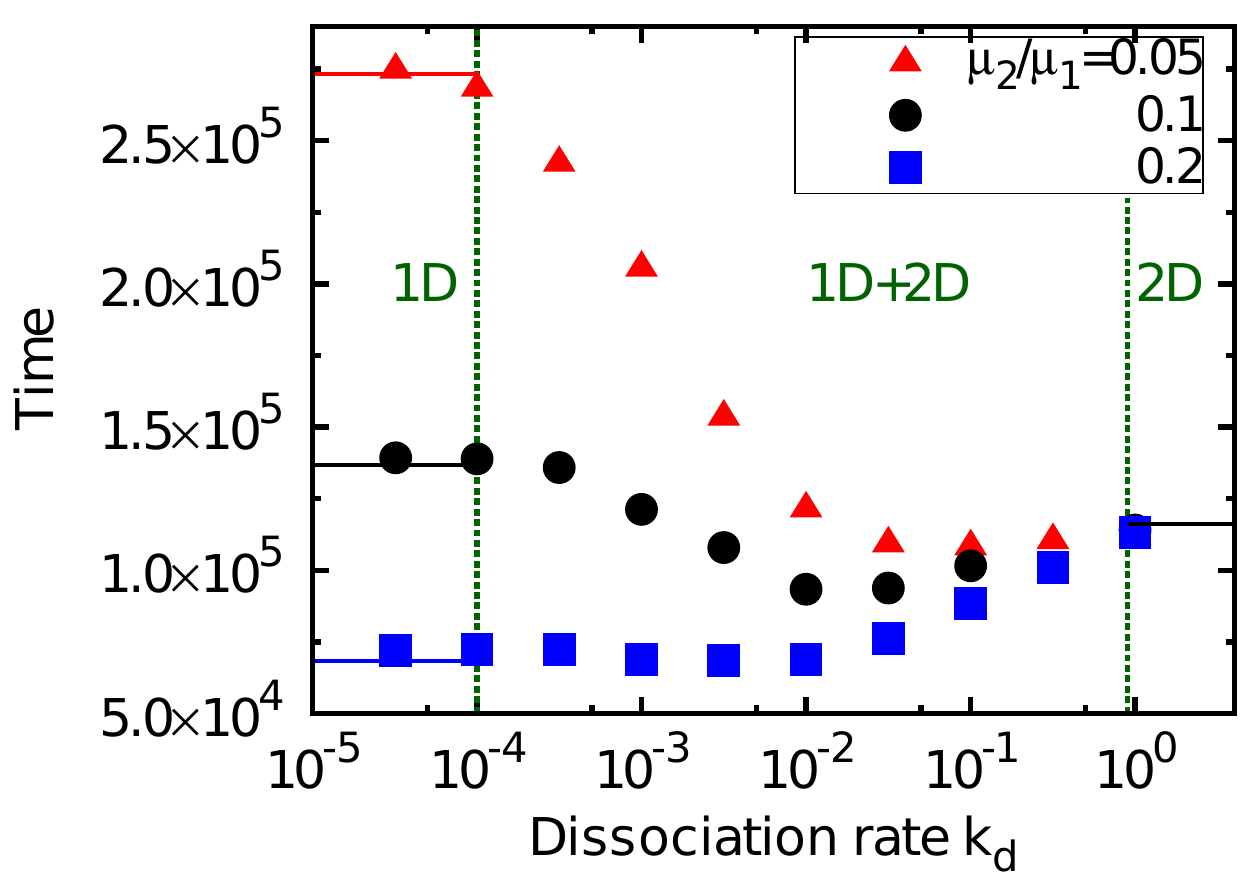}
    \caption{The target search time as a function of the dissociation rate $k_d$ for different sets of diffusion hopping rates. Here we take $L=100$. The two molecules are initially randomly distributed on the lattice with equal probability. We also set $k_{\text{off}}=0.25$. }
    \label{fig-4}
\end{figure}

\begin{figure*}
    \centering
    \includegraphics[width=0.6\columnwidth]{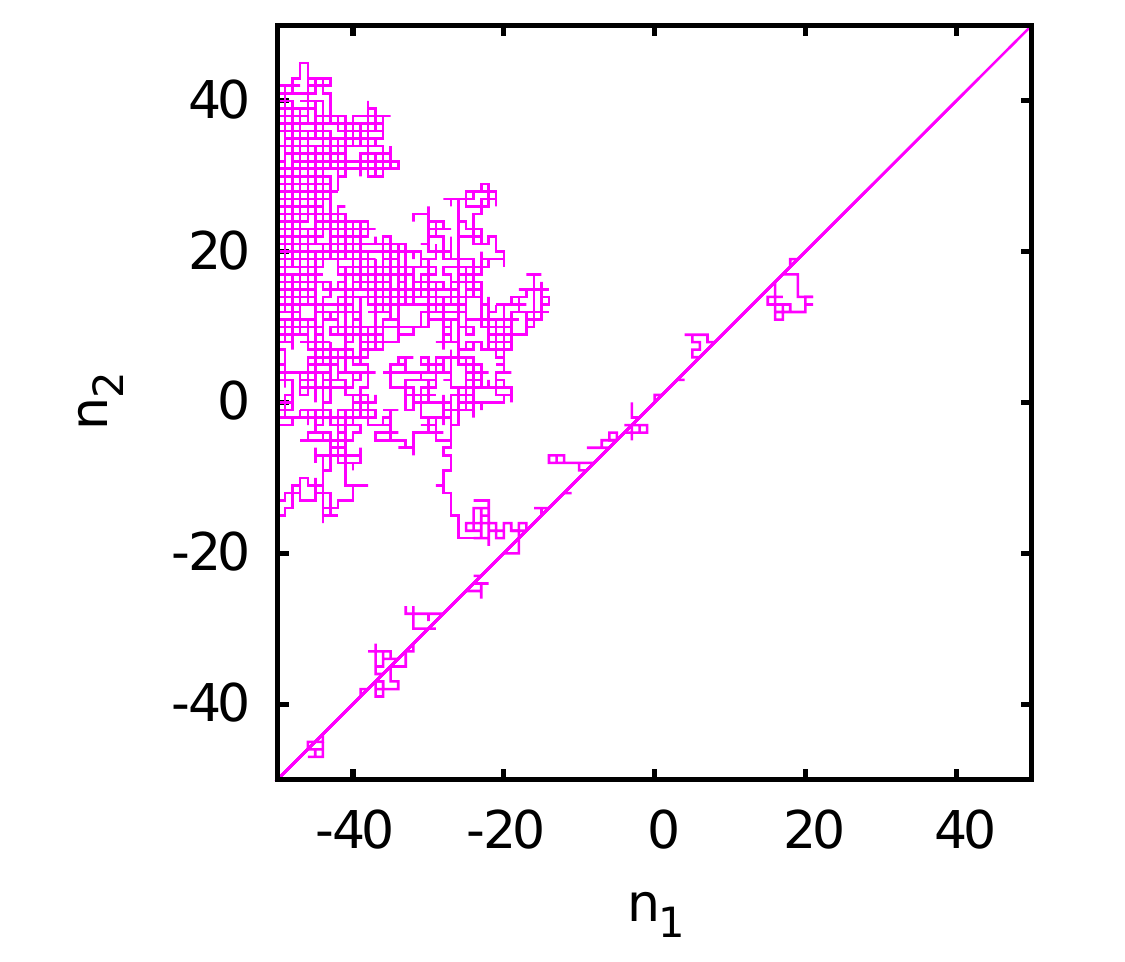}
    \includegraphics[width=0.6\columnwidth]{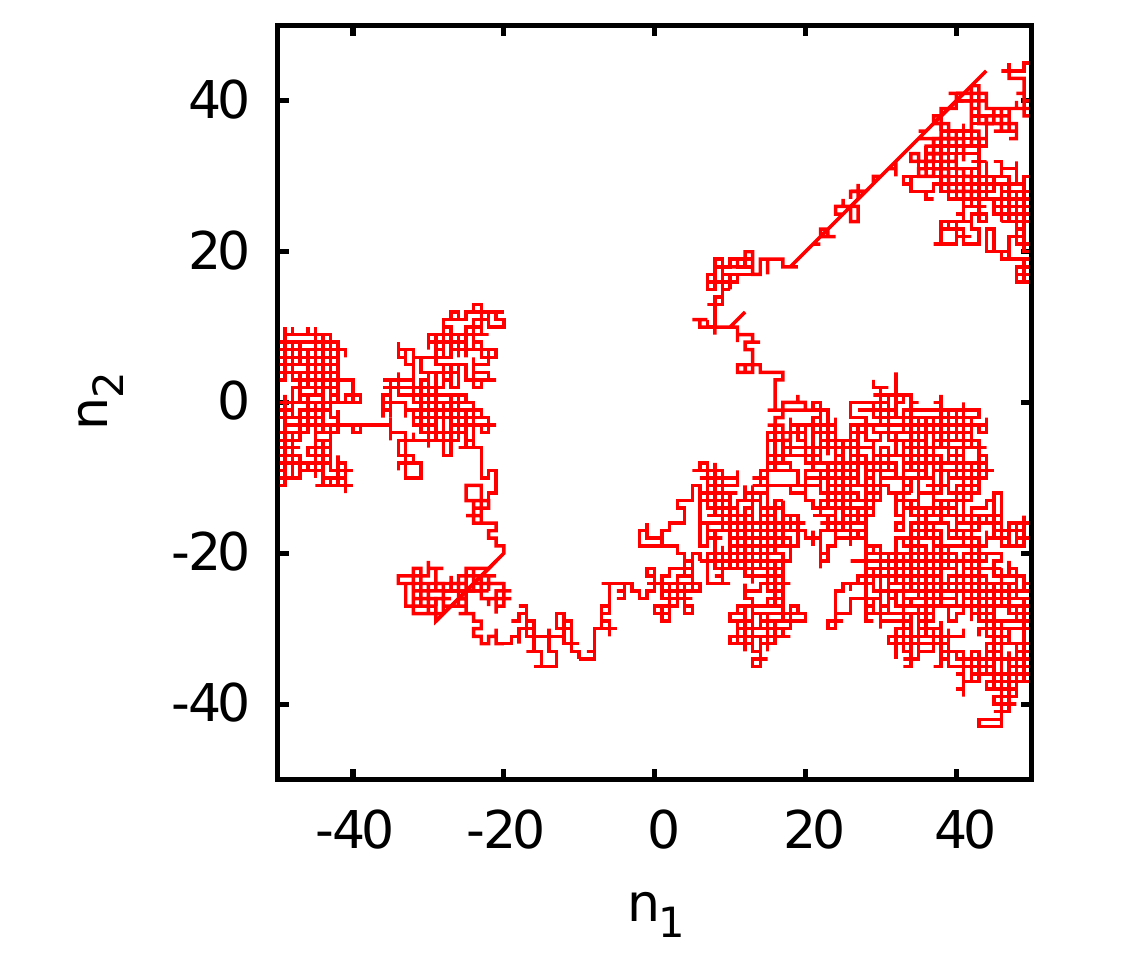}
    \includegraphics[width=0.6\columnwidth]{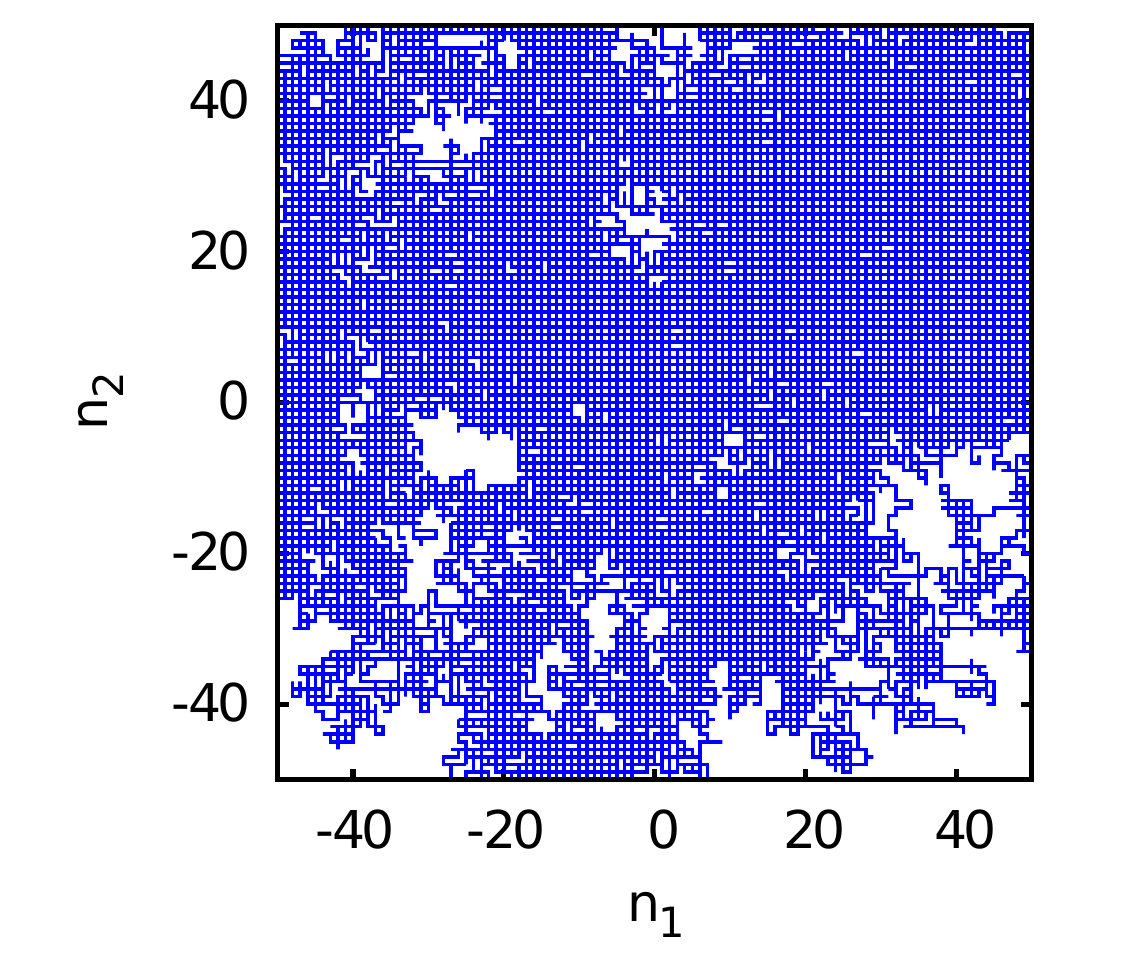}
    \caption{The trajectories of the searching molecules. Here the lattice size $L=100$ and the initial position of two molecules are $n_1=-50$ and $n_2=0$.  The target is at the end of the interval, i.e., at $(L/2,L/2)$. For simulations the following parameters were utilized:  $\mu_2 / \mu_1=0.1$, $k_{\text{off}}=0.25$, and three different values of $k_d= \{10^{-4}, 10^{-2}, 1 \}$  for left, middle and right panels, respectively.}
    \label{fig-5}
\end{figure*}

To better understand this dynamic behavior, we analyze the trajectories of the searching particles in the system by tracking the positions of the two molecules ($n_1$, $n_2$) for different values of the dissociation rate $k_d$. The corresponding trajectories are shown in  Fig. \ref{fig-5}. For the strong inter-molecular attractions (small $k_d$), once two molecules form the complex by reaching the same site, the complex moves as a one particle with rare dissociations (Fig. \ref{fig-5}, left). In this case, one can clearly see the motion along the diagonal line, emphasizing the 1D nature of this search regime. For strong inter-molecular repulsions (large $k_d$), the two molecules move independently at all times. The 2D space is almost fully explored in this regime (Fig \ref{fig-5}, right). At the intermediate values of inter-molecular interactions, the molecules move both as the complex and as the monomers (see Fig. \ref{fig-5}, middle). When the complex is formed, the trajectory follows only along the diagonal direction in the ($n_1, n_2$) coordinates. But these diagonal excursions are typically short because the hopping rate $\mu_2$  is slow. When the molecules move independently as two particles, they can move faster, which should  accelerate the search.  But the particles also explore the space, and this slows down the overall search because regions far away from the target probed repeatedly. However, for the intermediate values of $k_d$, the optimal balance between these two trends  might be found. The overall search time might be low because the target can be found from both the 1D mechanism (coming from the diagonal) and from the 2D mechanism(coming from all other directions). This phenomenon is similar to the protein facilitated diffusion when the faster dynamics is found for the regime that combines 1D and 3D motions  \cite{sheinman2012classes,benichou2011, mirny2009,veksler2013speed,shvets2018mechanisms}.

\begin{figure}
    \centering
    \includegraphics[width=0.9\columnwidth]{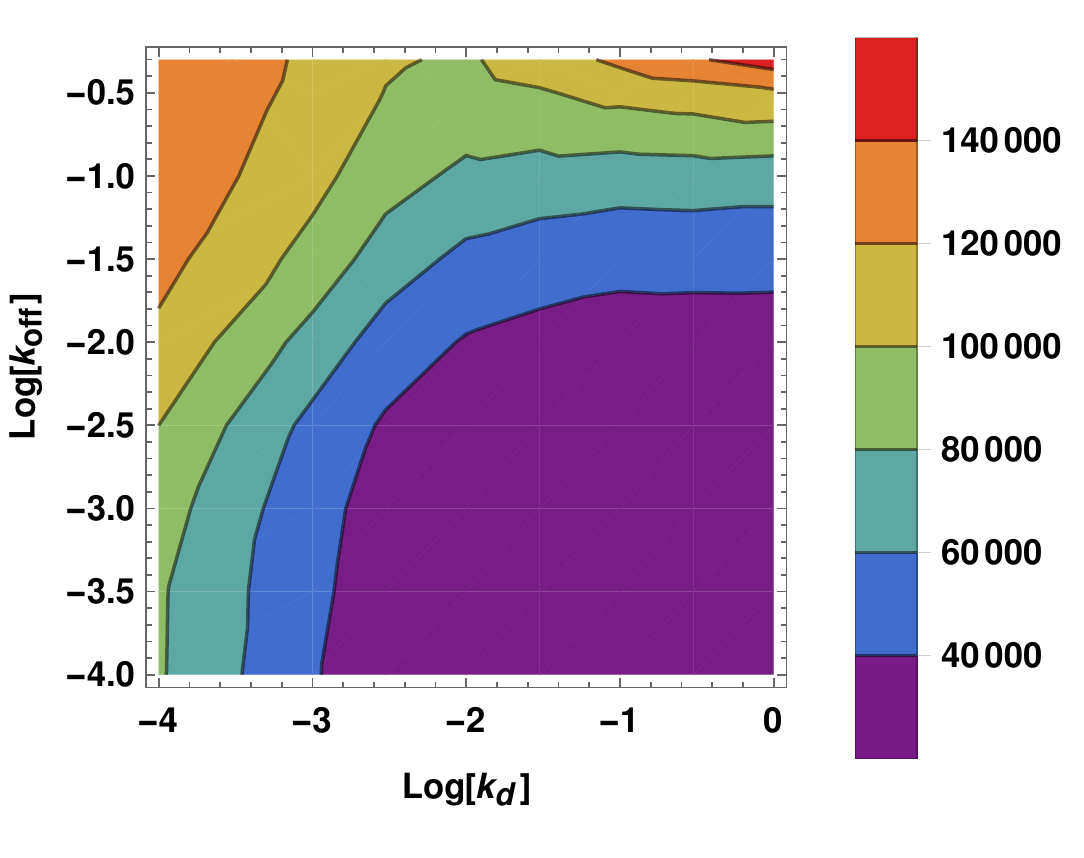}
    \caption{Target search times as the function of the dissociation rate $k_d$ and the unbinding rate $k_{\text{off}}$. For simulations the following parameters were used: $L=100$ and $\mu_2 / \mu_1=0.1$. The target is at the end of the interval.}
    \label{fig-6}
\end{figure}

The target search dynamics shows a rich behavior, and we identified the target residence time ($1/k_{\text{off}}$) and the strength of interactions (in terms of the dissociation rate $k_{d}$) as main factors controlling the system. One can see this from Fig. \ref{fig-6}, which shows the heat map plot of the MFPT as functions of the kinetic rates $k_{off}$ and $k_{d}$. The fastest target search is observed for strong repulsions and large target residence times (large $k_{d}$ and small $k_{\text{off}}$): see  Fig. \ref{fig-6}.  In this case, the system does not form the slowly moving complex and keeping one particle at the target site for a long time decreases the 2D space exploration. The slowest target search is predicted for short residence times and strong inter-particle attractions (large $k_{\text{off}}$ and small $k_{d}$). These conditions favor the formation of the complex, which moves very slowly. The non-monotonic behavior of the MFPT as the function of the dissociation rate $k_{d}$ is observed for short residence times (large $k_{\text{off}}$). Here,by tuning the interaction strength, it is possible to reach the target in the most optimal fashion by combining the effectively 1D and 2D mechanisms.

Our theoretical method allows us to explore the role of the target location on the search dynamics. The results are presented in Fig. \ref{fig-m} for different values of the dissociation rate $k_{d}$. As expected, the search is faster for target in the middle of the interval in comparison with the target location at the end of the interval. This is because the average distance between the starting locations of the two particles (uniformly distributed along the interval) are shorter when the target is inside the segment.  For small dissociation rates, the target search is effectively 1D process because the complex is the dominating state in the system. It can be shown then that for the target at the site $m$ \cite{veksler2013speed},
\begin{equation}
T_{1D}(m)=\frac{L^2+2L+12 m^2}{12 \mu_2}.
\end{equation}
This result agrees reasonably well with computer simulations: compare red symbols and a red curve. For the set of parameters in Fig. \ref{fig-m}, the mean search time is lowest for the intermediate value of $k_d$ irrespective of the target location. 

\begin{figure}
    \centering
    \includegraphics[width=0.9\columnwidth]{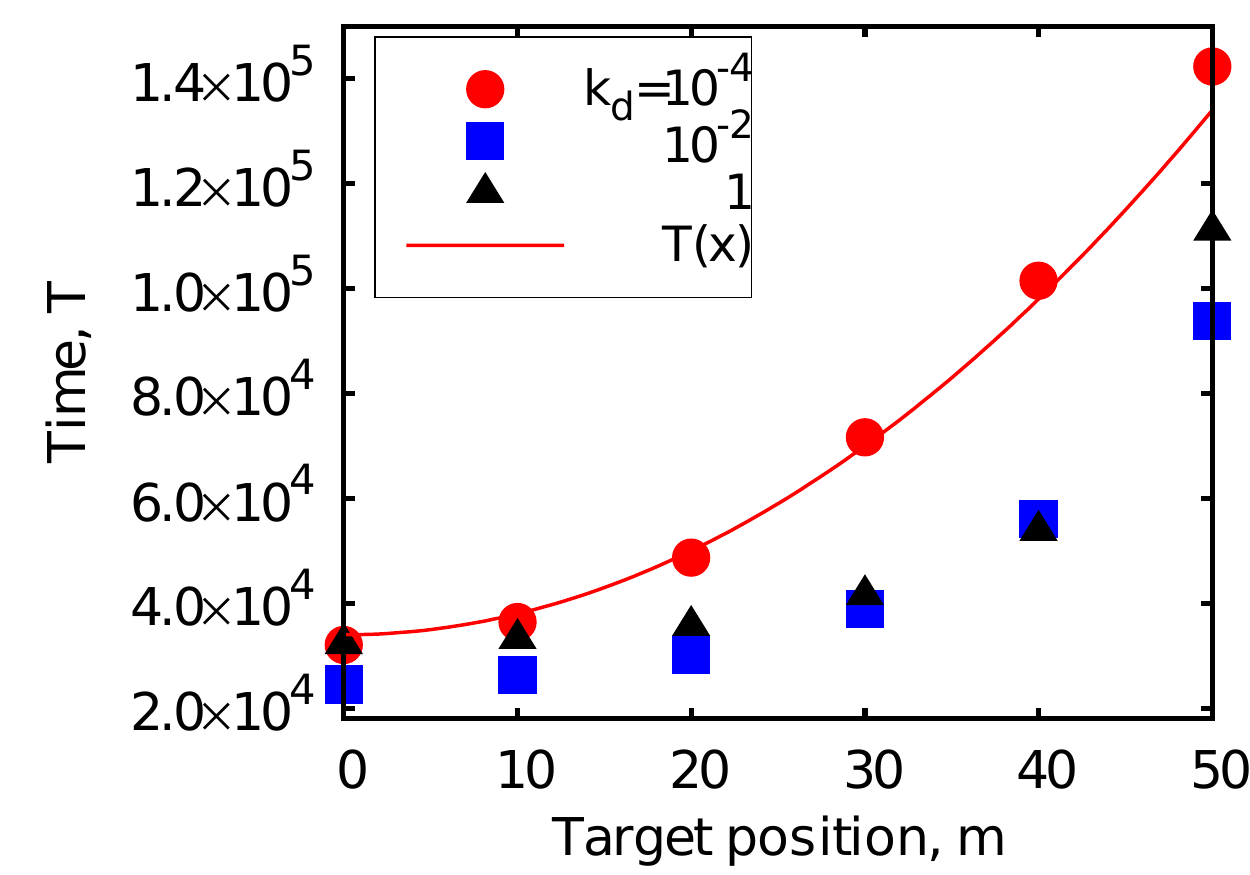}
    \caption{Search time as a function of the target position $m$ for three different values of $k_d$. Here the lattice size $L=100$, $\mu_2/\mu_1=0.1$.}
    \label{fig-m}
\end{figure}

\section{Summary and Conclusions}
\label{sec-conclusion}
Protein search processes for specific sites on DNA have been intensively studied theoretically in recent years, but the interactions between participating molecules have been mostly ignored. To fill this theoretical gap, we developed a minimal theoretical model of the target search by two molecules that are exploring 1D segment of sites, one of which is the target. In our model, the molecules that occupy the same binding site can reversibly transition into the dimer complex, and these association/dissociation processes reflect the interactions between them.  The dimer complex can also move along the lattice, although slower than the individual particles. The molecules have finite residence times at the target sites, and the search is accomplished when both of them are found at the target simultaneously. 

We investigated the target search using analytical calculations and Monte Carlo computer simulations. Our theoretical analysis shows that three dynamic search regimes are possible in the system depending on the strength of inter-molecular interactions. For strong attractive inter-molecular interactions, the particles quickly transition into the dimer complex, which performs the search for the target. This dynamic regime is effectively one-dimensional. In the opposite limit of strong repulsions, the dimer complex cannot exist in the system and the two molecules perform the search for the target. This regime is effectively two-dimensional because of independence of searching molecules. For the intermediate range of interactions, the search regime can be identified as 1D+2D because the system switches between the dimer complex and the two independent particles. It is also found that there might be the optimal strength of interactions at which the dynamics is the fastest due to combining two possible mechanisms of the target search (1D+2D). In addition, our calculations show that increasing the target residence time of the particles should accelerate the search. Furthermore, the search dynamics is faster for targets far away from the boundaries of the system. 

Although our theoretical method clarifies many aspects of the inter-molecular interactions in the target search processes, one should notice that the approach is rather oversimplified with many realistic features not taken into account. More than two transcription factor proteins are typically involved in gene activation or repression \cite{allemand2006}. We considered only the one-dimensional motion of the particles, which corresponds to the protein translocation along the DNA segment. However, in real cells transcription factors alternate between scanning the DNA chain and moving in the bulk solution around DNA. In addition, the sequence specificity of the DNA segments and the protein conformational fluctuations are totally neglected, while theoretical studies suggest that these factors might strongly influence the search dynamics \cite{shvets2018mechanisms}. It will be important to test the presented theoretical ideas using more advanced theoretical analysis as well as in the experimental studies.

\section*{Acknowledgments}

The work was supported by the Welch Foundation (Grant C-1559), by the NSF (Grant CHE-1360979), and by the Center for Theoretical Biological Physics sponsored by the NSF (Grant PHY-1427654).

\bibliography{rsc} 

\end{document}